\documentclass[12pt]{article}
\usepackage{a4wide,epsfig}

\def\det{{\rm det}}
\def\eg{{\em e.g.}}

\def\mev{\,{\rm MeV}}
\def\mfp{\,{\em mfp}}

\def\erg{\,{\rm erg}}
\def\ie{{\em i.e.}}
\def\km{\,{\rm km}}
\def\cm{\,{\rm cm}}

\def\s{\,{\rm s}}
\def\K{\,{\rm K}}
\def\tr{{\rm Tr}}
\def\wrt{{\it w.r.t.}}

\def\grho{g_{\rho\pi\pi}}
\def\gb{Goldstone bosons}

\newcommand{\nn}{\nonumber}
\newcommand{\be}{\begin{equation}}
\newcommand{\ee}{\end{equation}}
\newcommand{\ba}{\begin{eqnarray}}
\newcommand{\ea}{\end{eqnarray}}

\newcommand{\morder}[1]{{\cal O}\left(#1 \right)}
\newcommand{\order}[1]{${\cal O}\left(#1 \right)$}
\newcommand{\wt}[1]{{\widetilde#1}}

\begin{document}

\thispagestyle{empty}
\begin{flushright}
NORDITA 2003-56 NP \\
hep-ph/0311342\\
\today
\end{flushright}

\begin{center}

\vskip \baselineskip
{\Large\bf Photon Emission from Dense Quark Matter} \\
\vskip 3\baselineskip

C.\ Vogt$^{a}$\,\footnote{Email: cvogt@nordita.dk},\
R.\ Rapp$^{a,b}$\,\footnote{Email: rapp@comp.tamu.edu}\
and R.\ Ouyed$^{c}$\,\footnote{Email: ouyed@phas.ucalgary.ca}
\vskip \baselineskip
 $^{a}${\small {\it Nordita, Blegdamsvej 17, 2100 Copenhagen, Denmark}}\\
 $^{b}${\small {\it Cyclotron Institute and Physics Department, 
      Texas A\&M University, College Station, TX 77843-3366, USA}}\\
 $^{c}${\small {\it Department of Physics and Astronomy, University of Calgary,
 2500 University Drive NW, Calgary, Alberta, T2N 1N4 Canada}}\\

\vskip 2\baselineskip

\end{center}

\vskip \baselineskip

\begin{abstract}
Thermal emission rates and mean free paths of photons in a
color-flavor locked (CFL) phase of quark matter at high densities
and moderate temperatures are evaluated. Our calculations are
based on a low-energy effective theory for CFL matter describing
Goldstone boson excitations and their electromagnetic as well as 
strong interactions. In-medium coupling strengths of vector mesons 
are estimated to be smaller than in vacuum.
As a consequence of in-medium modified pion dispersion relations,
novel processes such as $\pi^+ \pi^- \rightarrow \gamma$ and 
$\gamma \to \pi^+ \pi^-$ become possible. The total photon 
emissivity is found to be very large, exceeding contributions 
from thermal $e^+e^-$ annihilation above temperatures of about 5~\mev.
At the same time, the corresponding mean free paths become very small.
Our results imply that the photon flux from the surface of a 
(hypothetical) CFL star in its early hot stages saturates the 
black-body limit.  Schematic estimates for the early thermal evolution 
of the star are also presented.
\end{abstract}

\section{Introduction}

The discovery of asymptotic freedom, leading to the formulation of
quantum chromodynamics (QCD) as the theory of strong interactions,
was soon followed by the suggestion that matter at sufficiently
high densities consists of a deconfined phase of
quarks~\cite{col:1974}. Only shortly afterwards it was pointed
out~\cite{bar:1977} that the true ground state of cold dense quark
matter exhibits color superconductivity (CSC), characterized by
diquark condensation with an estimated energy gap $\Delta$ of the
order of 1~\mev\ between the highest occupied and the unperturbed 
quark state at the Fermi surface. Since this magnitude of the gap is 
rather small for phenomenological applications, CSC subsequently 
received little attention. The situation changed when
reinvestigations \cite{alf:1997,rapp:1997} using nonperturbative
forces (\eg, instanton-induced) showed that the gap can be
substantially larger, $\Delta\simeq100~\mev$ for moderate
quark chemical potentials, $\mu_q\simeq$~350~\mev. Similarly
large values are obtained from estimates based on perturbative
calculations at asymptotically high densities~\cite{pis:1998,son:1998}. 
Thus, from the practical point of view, the existence of color 
superconductivity in compact stars has (re-) emerged as an exciting 
possibility.

The detailed properties of CSC matter relevant to astrophysical
applications depend on the interplay of the quark chemical potential,
the $q$-$q$ interaction strength, and the bare masses of the (light)
quarks $u$, $d$ and $s$. In particular, for $\mu_q$ below the
(constituent) strange quark mass, only $u$ and $d$ quarks are
subject to BCS pairing. The corresponding phase is known as
2-flavor CSC (2SC). In the idealized case where the quark chemical
potential is much larger than the strange quark mass ($m_s$), the
latter becomes negligible and all three flavors exhibit likewise
pairing. The preferred symmetry (breaking) pattern in this phase
corresponds to the so-called color-flavor locking
(CFL)~\cite{alf:1998}, since the underlying diquark condensate is
invariant only under simultaneous color and flavor
transformations. In the present work, we will focus on the
CFL phase (for a recent review and a more exhaustive list of
references, cf.~\cite{sch:2003}).

The investigation of compact stellar objects largely relies on
electroweak emission spectra over a broad range of frequencies.
Thus, emissivities and mean free paths ({\em mfp}'s) of neutrinos and
photons are important quantities to assess bulk properties of matter
within compact stars.  It is therefore vital to address the
question in how far the existence of CSC inside compact stars affects
their properties, such as neutrino and photon emission as well as 
thermal evolution.
Pertinent studies have recently been carried out for CSC phases 
in various scenarios~\cite{jai:2002,red:2002,jai:2001}.
Ref.~\cite{jai:2002} mostly addressed neutrino emission for
CFL matter at temperatures in the sub-MeV range ($\sim$0.1~MeV
or $10^9$~Kelvin) relevant to the long-term evolution of
a compact star. In  Ref.~\cite{red:2002} the CFL neutrino problem
has been investigated for a temperature range of up to 30~\mev,
corresponding to the hottest phases expected to occur during a
supernova explosion or in the early stages of a proto-neutron 
star (see, \eg, Ref.~\cite{sha:1983}). Photon emission has been 
calculated in Ref.~\cite{jai:2001} in a weak-coupling expansion 
for (gapped) quarks. The objective of the present article is to 
evaluate both photon emissivities and \mfp's in the strong coupling 
regime of CFL matter in the (tens of) MeV temperature region making
use of an underlying effective theory description.
The main processes of interest therefore involve scattering and 
decay processes of \gb, taking into account their in-medium
modified dispersion relations.  For comparison, the effects of thermal 
electrons and positrons are assessed, which prevail towards 
lower temperatures. Furthermore, we incorporate generalized vector 
mesons, relevant at the high-temperature end of the CFL phase.
As a lower limit, and in order to minimize the number of poorly known 
parameters, such as meson masses, we restrict the effective theory
description to the $SU(2)$ flavor sector, \ie, neglect contributions from
mesons involving strange quarks (cf.~Ref.~\cite{Turbide:2003si} for a
recent analysis in hot hadronic matter). We will comment on the effects of
kaons and other strangeness-bearing mesons in the summary.

The outline of the article is as follows. In Sect.~\ref{sec:effL} we will
recall the basic elements of the effective low-energy Lagrangian which
describes the electromagnetic and strong interactions of pseudoscalar \gb.
In Sect.~\ref{sec:phot} the evaluation of photon emission rates and mean
free paths will be presented, including novel aspects specific to
the CFL setting. Sect.~\ref{sec:evo} is devoted to an estimate of the
photon-driven cooling behaviour in the early stages of the thermal 
evolution of a compact star which accommodates CFL matter throughout. 
Finally, Sect.~\ref{sec:sum} contains our summary and conclusions. 
Explicit expressions for medium-modified Feynman rules and process 
amplitudes have been relegated to the Appendix.

\section{Effective Low Energy Lagrangian}
\label{sec:effL}

The effective low-energy theory of CFL diquark matter is
analogous to the chiral theory in the QCD  vacuum and has been
introduced in Refs.~\cite{hong:1999,cas:1999,son:1999}. Most
notably, color-flavor locking induces the breaking of chiral and
baryon number symmetries. Thus, an octet and singlet of pseudoscalar 
\gb\ emerges which constitute the low-energy excitations and hence
determine the features of CFL matter for energies below the gap, 
including electromagnetic and thermal properties.  Although
the (pseudo-) \gb\ in CFL matter carry the same quantum numbers as
in vacuum, differences in the chiral symmetry breaking pattern
render their masses -- arising form the explicit breaking due to
finite current masses -- qualitatively and quantitatively different.
While in the vacuum case the parameters of the chiral Lagrangian
have to be extracted from experiment, in the case of CFL matter a
matching of the low-energy sector with the so-called high-density
effective theory (HDET) allows for a determination of parameters
in lowest order of the chiral and perturbative expansion,
respectively. A detailed analysis for excitations below and above 
the gap revealed, \eg, that the mass hierarchy in the strange and 
nonstrange sector is (partially) 
inverted~\cite{son:1999,bed:2001,sch:2001}.

The main objective of the present work is an assessment of
electromagnetic and strong processes involving Goldstone-boson and 
vector excitations. As a convenient approach towards this goal, we 
here adopt the hidden local symmetry (HLS) framework that, in the context
of CFL matter, was suggested in Ref.~\cite{rho:2000} and further elaborated 
in Ref.~\cite{jai:2001}.  It amounts to a treatment 
of vector mesons as dynamical, composite gauge bosons corresponding 
to a hidden local $SU(3)_{c+L+R}$ symmetry.\footnote{Strictly speaking, 
hidden local symmetry only holds in the zero size approximation, 
neglecting the finite size of condensed diquarks, 
cf.~Ref.~\cite{rho:2000} for a discussion of this issue.}  
Originally, the HLS approach was introduced in the vacuum case in 
Ref.~\cite{ban:1985} (for a review see~\cite{ban:1987}). 
Here, we will briefly recall the main elements important for our work. 

We start with the effective chiral Lagrangian of (generalized) 
pseudoscalar \gb\ in the CFL phase, which  is equivalent to the 
non-linear sigma model known from the vacuum and 
reads~\cite{hong:1999,cas:1999,son:1999}  
\ba
 {\cal L}_{\rm eff}=\frac{f_\pi^2}{4} \, 
 \tr \, [ \partial_0 U \, \partial_0 U^\dagger-v_\pi^2\, 
       \partial_i U \, \partial_i U^\dagger ] 
   -c \,[\det (M) \, \tr \, (M^{-1} U) + h.c. ]  \  ,
\label{eq:chilgn}
\ea
with $M={\rm diag} (m_u,m_d,m_s)$ being the current quark mass matrix.
The chiral field $U$ is related to the Goldstone boson octet 
$\pi \equiv \pi^a \lambda^a$ by  
\ba
 U = e^{i\,\pi / f_\pi} \,,
\label{eq:U}
\ea
where the Gell-Mann flavor matrices are normalized as 
$\tr [\lambda^a,\lambda^b]=2 \, \delta^{ab}$ and $f_\pi$ is the 
(in-medium) pion decay constant. The latter, as well as the constant 
$c$ in Eq.~(\ref{eq:chilgn}), can be obtained from the matching procedure, 
referred to above, to leading order in $\alpha_s$~\cite{son:1999},  
\ba
f_\pi^2 = \frac{21- 8 \, \ln 2}{18} \frac{\mu_q^2}{2 \pi^2} \,,
\qquad
c = \frac{3\Delta^2}{2\pi^2} \,.
\ea
The explicit expressions for the pion masses then take the 
form~\cite{sch:2001}
\ba
m_{\pi^\pm} &=& \mp \frac{m_d^2-m_u^2}{2 \mu_q} 
 + \bigg[ \frac{3 \Delta^2}{\pi^2 f_\pi^2} \, (m_u+m_d) \, m_s \bigg]^{1/2} \ .
\label{eq:mpi}
\ea

As Lorentz invariance is broken due to the rest frame of the medium, 
the pion decay constant (defined via the weak pion decay matrix element)  
generally splits into temporal and spatial 
components~\cite{pis:1996}, denoted by $f_T$ and $f_S$, respectively.
They are related to the pion velocity $v_\pi$ (found to be 
$v_\pi=1/\sqrt{3}$, coinciding with the speed of sound in 
relativistic fluids~\cite{son:1999}) through $f_S=v_\pi^2 f_T$, 
with $f_T=f_\pi$. Consequently, the pion velocity also appears in 
the effective Lagrangian~(\ref{eq:chilgn}), and 
induces a modified dispersion relation for the \gb, which follows 
from the corresponding kinetic term as 
\ba
E = \sqrt{m_{\rm GB}^2 + v_\pi^2 \, {\bf p}^2} \, .
\ea

The hidden local symmetry of the low-energy effective theory, 
Eq.~(\ref{eq:chilgn}), can be made explicit by decomposing the 
left and right pion fields $\pi_{L,R}$ according to 
\ba 
 U_{L,R} = e^{i\,\pi_{L,R} / f_\pi} = \xi^T_c \, \xi_{L,R} \ ,
\ea
where $\xi_c$ and $\xi_{L,R}$ are $SU(3)$ color and chiral gauge
fields, respectively. In the following, we will fix the local 
$SU(3)_{c+L+R}$ gauge by $\xi_c = \xi_{L,R} \equiv \xi$, which 
identifies the left and right pion fields and reflects 
the color-flavor locking property. In terms of the field $\xi$, 
the kinetic term of the Lagrangian~(\ref{eq:chilgn}) then reads
\ba 
 {\cal L}_{\rm kin} = -\frac{f_\pi^2}{4} \, 
 \tr \, \Big[ \partial_0 \xi \, \xi^\dagger - \xi \, \partial_0 \xi^\dagger 
    + v_\pi^2 \, (\partial_i \xi \, \xi^\dagger 
    - \xi \, \partial_i \xi^\dagger ) \Big]^2 \ .
\label{eq:kinlgn}
\ea
We proceed by introducing both the vector field, 
$V_\mu = V_\mu^a \lambda^a / 2$, and the in-medium photon 
field, $\wt{A}_\mu$, by means of the covariant derivative~\cite{ban:1985} 
\ba 
D_\mu \xi = \partial_\mu \xi -i \, \wt{g} \, V_\mu \, \xi
            -i \, \wt{e} \, \xi \, \wt{A}_\mu \, Q \ ,
\label{eq:gauging}
\ea 
with $Q={\rm diag}(\, 2/3,-1/3,-1/3\, )$ being the quark-charge matrix.
The vector-meson field is rendered dynamical in the usual manner 
by introducing a kinetic term 
$ - (1/4) \, \tr\, F_{\mu\nu} F^{\mu\nu}$, 
where $F_{\mu\nu} = \partial_\mu V_\nu - \partial_\nu V_\mu
+ i \, \wt{g} \, [V_\mu,V_\nu]$ is the field strength tensor.

As noted in Ref.~\cite{alf:1998}, the electromagnetic
coupling in the CFL phase is modified.
Since a diquark condensate carries baryon number, the physical
(massless) photon field in medium is a linear combination of
the original photon and the eighth component of the gluon field.
The electric charge $\widetilde{e}$ in medium is therefore related
to the vacuum charge $e$ and strong coupling $g_s$ via
$\widetilde{e}= e\,\cos \theta$ where $\cos\theta = \sqrt{3} \,
g_{s}/\sqrt{3g_s^2 + 4\,e^2}$. Since, over a wide range of momenta,
the strong coupling is much larger than the electromagnetic one,
we will, for practical purposes, set $\cos\theta \simeq 1$ and drop
the tilde notation for the photon field and the electromagnetic 
coupling in the following.
The introduction of the electroweak interaction in the effective
Lagrangian of the CFL and 2SC phases is presented in detail in
Ref.~\cite{Casalbuoni:2000jn}.

The question now arises how to determine the the masses and 
coupling strengths of the vector mesons in the CFL environment.   
In Ref.~\cite{jai:2001}, where the HLS approach has first been adopted 
to evaluate dilepton emission rates from CFL matter, vector-meson dominance 
(VMD) was assumed in connection with $\wt{g}=g_{\rho\pi\pi}$ at its vacuum 
value ($\sim$~6). From the KSRF relation~\cite{KSRF} it was then inferred
that the vector-meson mass scales with the chemical potential $\mu_q$.
On the other hand, in Ref.~\cite{rho:2000} a detailed weak-coupling 
calculation based on a Bethe-Salpeter equation has
identified the axial/vector excitations (mesons) as diquark-hole
states resulting in
\be
 m_V= 2\, \Delta \, \sqrt{ 1-{\rm e}^{-C / \tilde{g}} } \ ,
 \label{eq:mrho}
\ee
which for all practical purposes equals twice the gap 
($C=(3-\sqrt{3}) \, \pi^2 \, \sqrt{6}\simeq 30.7$). 
In the recent work of Ref.~\cite{Jackson:2003dk}, 
the (parametric) dependence of Eq.(\ref{eq:mrho}),
$m_V\propto\Delta$,  was corroborated by a nonperturbative argument, 
\ie, by matching higher-derivative contact terms of the low-energy 
effective theory, Eq.~(\ref{eq:chilgn}), with local 
terms obtained as a result of integrating out the heavy $V$-fields 
in a generic chiral Lagrangian. Thus, we henceforth adopt 
$m_V = 2 \, \Delta$ also for the present case of a strong coupling 
environment, and do not distinguish between temporal 
and spatial vector-meson masses $m_T$ and $m_S$ as was done 
in Ref.~\cite{jai:2001}.
The vector-meson mass term in the Lagrangian then acquires the 
standard vacuum form $(m_V^2/2)\, V_\mu V^\mu$. 

Within the matching procedure of Ref.~\cite{Jackson:2003dk} it has 
furthermore been found that the vector-coupling strength scales 
according to\footnote{Using duality arguments,
the analysis in~\cite{Jackson:2003dk} also
exhibits that nucleon-type excitations are parametrically
heavy with their masses  proportional to $f^2_\pi/\Delta$. However,
at moderate densities, where $f_\pi$ is not larger
than the gap, this assertion no longer holds, and the (solitonic)
``nucleon'' or ``delta'' masses could become comparable to the
vector meson ones; for large
pairing gaps, it would thus be interesting to evaluate the effects
of $\pi$-$N$ interactions -- known to be strong in ordinary hadronic
matter -- in a CFL environment at temperatures above 10~MeV.}
\begin{eqnarray}
 \widetilde{g} \propto \frac{\Delta}{f_{\pi}} \ .
\label{eq:g-scaling}
\end{eqnarray}
This result suggests that, for large chemical potential, the 
self-interactions of the vector mesons are reduced with respect to 
(\wrt) the vacuum. By expanding the effective Lagrangian in the Goldstone 
boson fields, we see  that $\widetilde{g}$ is identical to the coupling 
strength $\grho$ of the vector meson to two pions. 
In particular, $\grho$ scales in the same way as $\widetilde{g}$. 
While in vacuum $\grho = \widetilde{g} \simeq 6$, the corresponding 
values in the CFL phase are parametrically suppressed. To obtain a more
quantitative estimate for the $\rho\pi\pi$ coupling, we follow  
Refs.~\cite{jai:2001,Jackson:2003dk} in assuming that the KSRF relations 
remain valid in CFL matter, \ie,  
\begin{eqnarray}
 m^2_V = 2 \, \widetilde{g}^2 f^2_{\pi} \ .
\label{eq:ksrf}
\end{eqnarray}
Here we have also assumed that VMD remains intact, 
implying that the HLS parameter has been fixed at $a=2$.\footnote{In  
the weak-coupling analysis of Ref.~\cite{man:2000} it was found that
the KSRF relations are modified at asymptotic densities, resulting in 
$a$=1. A similar feature arises within a renormalization-group
treatment of the HLS approach at the finite temperature chiral phase 
transition~\cite{har:2001}.}  
For the vector-meson interaction strength and their coupling 
to the \gb\ we then have\footnote{We would like to thank 
F.~Sannino for pointing this out to us.}
\begin{eqnarray}
 \grho = \widetilde{g} \simeq \sqrt2 \, \frac{\Delta}{f_{\pi}} \,.
\end{eqnarray}
With a typical parameter choice of $\Delta_0=$ 50 -- 150~\mev\ 
at $\mu_q\simeq 350~\mev $ (the latter implying $f_\pi\simeq 73~\mev$), 
one obtains $g_{\rho\pi\pi}\simeq$ 1 -- 2.8, substantially smaller than 
in free space.

For the reasons discussed above, we employ the vacuum form of the 
vector-meson mass term, implying that the vector-meson propagator is 
identical to the vacuum propagator. In order to preserve gauge 
invariance, it then turns out that also VMD needs to be applied in its 
vacuum version, \ie, without the distinction of temporal and spatial
components in the $V$-$\, \gamma$ vertex. Technically, this amounts to 
introducing into the Lagrangian a counter term of the form 
\ba
 4\, e\, \wt{g} \, f_\pi^2 \, 
 \tr \, ( A_0 \, Q \, V_0 - v_\pi^2 A_i \, Q \, V_i )
 - \frac{2\, m^2_V}{\wt{g}} \, e \, A_\mu \, \tr \, ( Q \, V^\mu ) \ ,
\ea
where the last term represents the standard $V$-$\, \gamma$ 
vertex\footnote{A similar procedure of restoring gauge invariance is 
often adopted when, \eg, including vertex form factors in hadronic 
effective theories~\cite{kap:1991}. The underlying microscopic origin of
the counter terms remains open.}. 
Thus, after substituting the covariant derivative~(\ref{eq:gauging}) in 
expression~(\ref{eq:kinlgn}) and expanding in the pion field $\pi$, the 
complete interaction part of our Lagrangian for \gb\ now reads
\ba
 {\cal L}_{\rm GB, int} &=& 
  - i \, \frac{\grho}{2} \, \tr \, \Big( V_0 \, [ \pi , \partial_0 \pi ]  
  + v_\pi^2 \, V_i \, [ \pi , \partial_i \pi ] \Big) 
  + \frac{e^2}{4} ( A_0^2 - v_\pi^2\, A_i^2) \, \tr \, ( \, [ Q , \pi ] \, )^2
  \nn \\ 
  && - \frac{e\, \grho}{2} \, \tr\, \Big( A_0 \, [ V_0, \pi ] \, [ Q , \pi ] 
  - v_\pi^2 \, A_i \, [ V_i, \pi ] \, [ Q , \pi ] \Big)
  - \frac{2\, m^2_V}{\wt{g}}\, e\, A_\mu\, \tr\, ( Q\,V^\mu )  \ .
\label{eq:intlgn}
\ea
VMD reflects itself by the absence of a direct $\pi\pi\gamma$ 
interaction term.
Finally, the physical photon and vector-meson field can be obtained 
by the usual diagonali\-zation procedure. Note that, within the present 
approach, all parameters of the low-energy effective Lagrangian are now 
fixed.

\section{Photon Emission and Absorption}
\label{sec:phot}

Let us turn to the evaluation of the photon emission rates and mean 
free paths in CFL matter as a function of temperature and  
the value of the zero-temperature gap. 
As mentioned in the introduction, we restrict ourselves to
processes within the $SU(2)$ flavor sector, \ie, charged and neutral 
pions, as well as rho mesons. Since at temperatures of
\order{1 \mev}, $e^+ e^-$ pairs have comparatively large number
densities~\cite{red:2002}, they will also be included in our analysis.
For definiteness, we assume the validity of the BCS relation
for the gap, $\Delta = \Delta_0 \, \sqrt{1-(T/T_c)^2}$, with a
critical temperature  $T_c \simeq 0.56 \, \Delta_0$.
In order to demonstrate the dependence of our results on the 
(poorly known) gap, we perform our calculation with two 
different values, $\Delta_0 = 50 \mev$ and $\Delta_0 = 150 \mev$.
Unless otherwise stated, we set $\mu_q = 350 \mev$.  

\subsection{Emission Rates and Emissivities}

The general set-up for the calculation of photon emission rates
follows the ones performed in an ordinary hot meson gas in
Refs.~\cite{kap:1991,song:1993}. The obvious differences expected in a
CFL phase are due to the different masses and strong couplings of the 
pions and rho mesons, as well as the modified pion dispersion relation.
Both of these features lead to an increase in available phase space.
The CFL pions are considerably lighter than in vacuum. Fixing the current
quark masses at  $m_u = 3~\mev$, $m_d = 7~\mev$ and $m_s = 120~\mev$,
we find $m_\pi \simeq 12$ (38)~\mev\ for $\Delta_0 = 50$ (150)~\mev\
and temperatures of \order{10 \mev}. Throughout our numerical calculations, 
the temperature dependence of the pion mass, entering through $\Delta(T)$ 
in Eq.~(\ref{eq:mpi}), will be accounted for.

The relevant processes which lead to one or two photons in the final
state, are scattering of the form $M_1 \, M_2 \to M_3 \, \gamma$ 
(where $M_i$ are pseudoscalar or vector mesons), annihilation,
$M_1 \, M_2 \to \gamma \, \gamma$, and rho-meson decays, 
$\rho \to \pi \, \pi \, \gamma$.
The (anomalous) neutral pion decay, $\pi^0 \to \gamma \, \gamma$, has
been considered in Ref.~\cite{jai:2002} for low temperatures of
\order{0.1 \mev}, where the emissivity was found to be exponentially
suppressed. For completeness, we have repeated the calculation for the
larger temperatures we are interested in here. Note that purely
electromagnetic processes, \eg, $\pi^+ \pi^- \to \gamma \, \gamma$,
are relatively suppressed by a factor $e^2/\grho^2\simeq$ 10 -- 100 as 
compared to strong processes such as $\pi^+ \pi^- \to \rho^0 \, \gamma$.
At low temperatures, however, strong processes, requiring at least one
in- or outgoing rho meson, are penalized by thermal suppression
due to the relatively large rho mass. The in-medium Feynman rules 
and explicit expressions for the amplitudes are summarized
in Appendix~\ref{app_feynman} and \ref{app_amplitude}, respectively.
We have verified that all amplitudes satisfy electromagnetic gauge 
invariance.

In addition to the above processes, which can all occur in a vacuum
environment, a novel photon source arises due to the medium-modified 
pion dispersion relation. With the in-medium pion velocity $v_\pi<$~1,    
annihilation of the type $\pi^+ \pi^- \to \gamma$ can become 
kinematically allowed, \ie, with all particles being {\em on-mass shell}.  
This is rendered possible due to the pion invariant-mass, 
$p^2 = m_\pi^2 - (1-v_\pi^2) {\bf p}^2$, being reduced and
eventually turning spacelike at large 3-momentum. 
Thus, sufficiently fast pions can annihilate on slow ones. 
A more detailed discussion of this effect, which also proceeds in
the inverse direction, thereby contributing to photon absorption, 
will be given in Sect.~\ref{sec:lbreaking}.

The photon emission rate for a given process, \eg, 
$M_1(p_1) \, M_2(p_2) \to M_3(p_3) \, \gamma(q)$, is obtained from 
a phase space integral over the pertinent unpolarised scattering 
amplitude, ${\cal M}$, according to  
\ba 
\label{eq:rate}
 E \frac{dR}{d^3 q} &=& \frac{\cal N}{2\, (2 \pi)^3}
 \int\frac{d^3 {\bf p}_1}{(2\pi)^3 \, 2 E_1}
 \int\frac{d^3 {\bf p}_2}{(2\pi)^3 \, 2 E_2}
 \int\frac{d^3 {\bf p}_3}{(2\pi)^3 \, 2 E_3} \,
 (2\pi)^4 \, \delta^{(4)}(p_1+p_2-p_3-q) \nn\\
 &&\times \Big| {\cal M}^{M_1 M_2 \to M_3 \gamma} \Big|^2 \, 
 f(E_1,T) \, f(E_2,T) \, [1 + f(E_3,T)] \,,
\ea
where ${\cal N}$ is a (spin-isospin) degeneracy factor and
$f(E_i,T)$ are Bose-Einstein distribution functions.
To evaluate the integral, we rewrite
\ba
 \int\frac{d^3 {\bf p}_i}{(2\pi)^3 \, 2 E_i} =
 \int\frac{d^4 p_i}{(2\pi)^3}\,\delta(E_i^2-m_i^2-v_i^2 \, {\bf p}_i^2)
 \, \Theta(E_i) \,,
\ea
with $v_i$ being the velocity of particle $i$ (different from 1 only for
\gb\ in the present work). One can then perform one of the 4-dimensional
phase space integrations (for definiteness, $i=3$) exploiting the 
$\delta$-function in Eq.~(\ref{eq:rate}). Since rotational symmetry 
persists in the medium, we may fix
the photon momentum ${\bf q}$ in $z$-direction for subsequent purposes.
The remaining angular dependences consist of the cosine of the polar 
angles, $\cos\theta_1$ and $\cos\theta_2$, of particles 1
and 2 (\wrt~the $z$-axis), and their relative azimuth
$\phi=\phi_1-\phi_2$ (the integration over the total azimuth
$\Phi=\phi_1+\phi_2$ becomes trivial).
With $d^3 p_j = |{\bf p}_j|^2 \, d \, |{\bf p}_j| \, d \, \Omega_j$
and $E_j \, d E_j = v_j^2 \, |{\bf p}_j| \, d \, |{\bf p}_j|$
for $j$=1, 2, we arrive at
\ba
 E \frac{dR}{d^3 q} &=& \frac{2 \pi\, {\cal N}}{8\, (2 \pi)^8}
  \int\frac{|{\bf p}_1| \, d E_1 \, d\cos\theta_1}{v_1^2}
  \int\frac{|{\bf p}_2| \, d E_2 \, d\cos\theta_2}{v_2^2} \, d\phi \; 
  \Theta(E_1+E_2-E) \nn \\
  &&\times \delta\Big( (E_1+E_2-E)^2-m_i^2-v_i^2 \,
           ({\bf p}_1+{\bf p}_2-{\bf q})^2 \Big) \, 
    \Big| {\cal M}^{M_1 M_2 \to M_3 \gamma} \Big|^2_{p_3=p_1+p_2-q} \nn\\
  &&\times \, f(E_1,T) \, f(E_2,T) \, [1+f(E_1+E_2-E,T)]  \,.  
\ea
For external (on-shell) particles, the energy integrations
are restricted from below by $E_{i,{\rm min}}=m_i$. The $\phi$
integration can be performed using the remaining $\delta$-function
at its zero argument, $\phi_0$,  generating
a Jacobian factor $[2\, v_i^2\, |{\bf p}_1| \, |{\bf p}_2| \,
| \sin\theta_1 \, \sin\theta_2\, \sin\phi_0|]^{-1}$. 
The last four integrations are carried out numerically.
\begin{figure}[!tb]
\begin{center}
\psfig{file=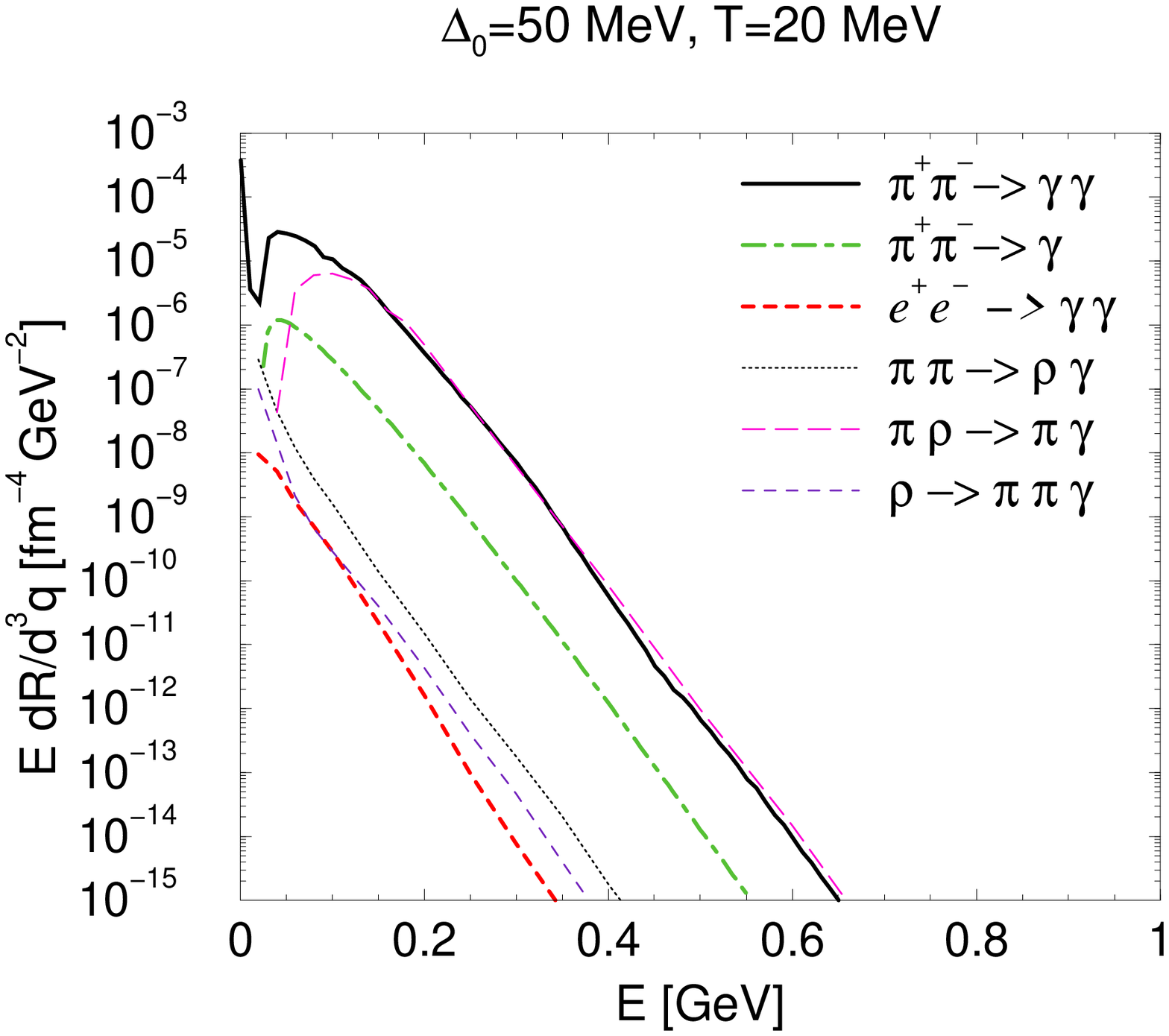,width=0.49\textwidth}
\psfig{file=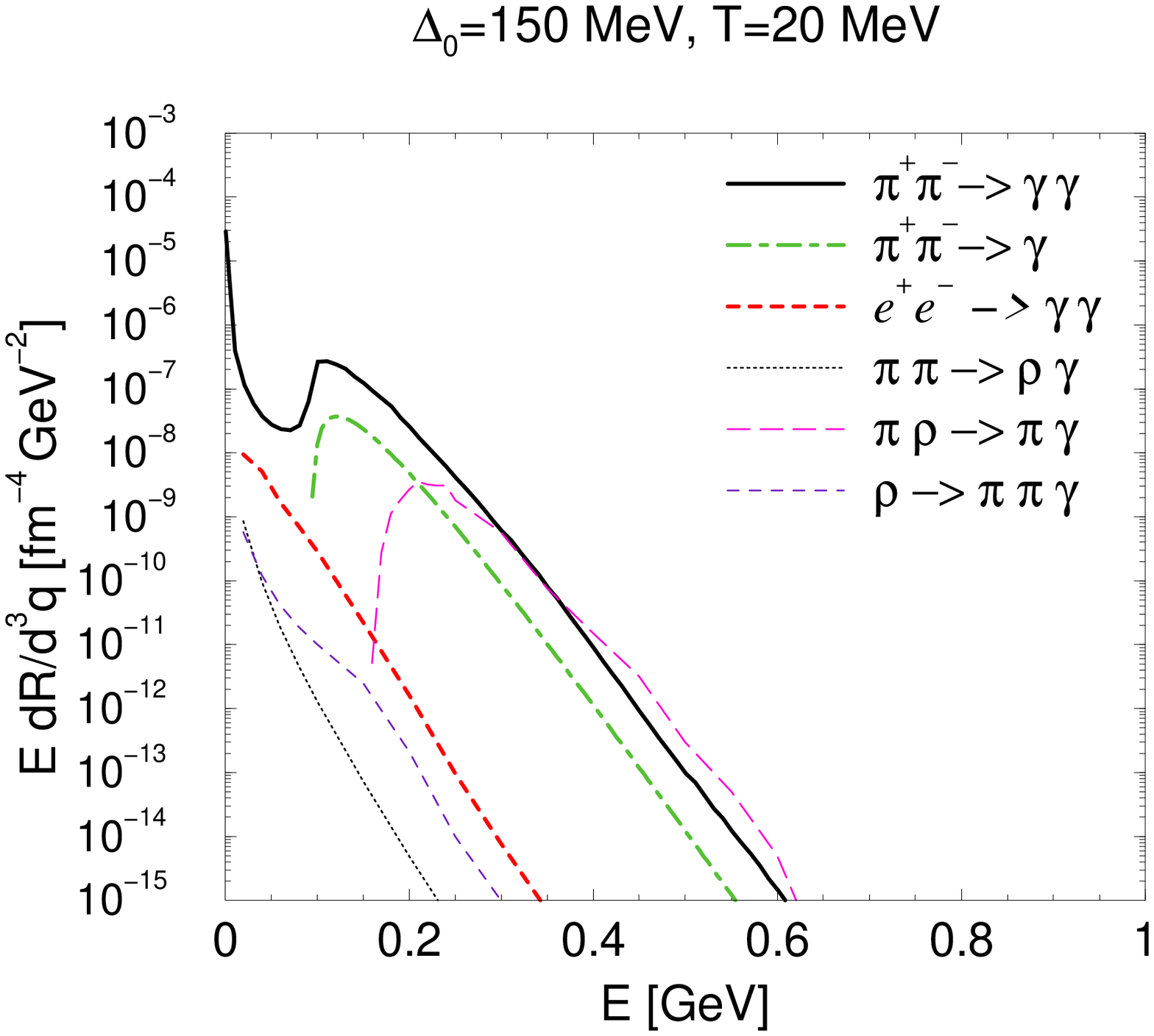,width=0.49\textwidth}

\vspace{0.5cm}

\psfig{file=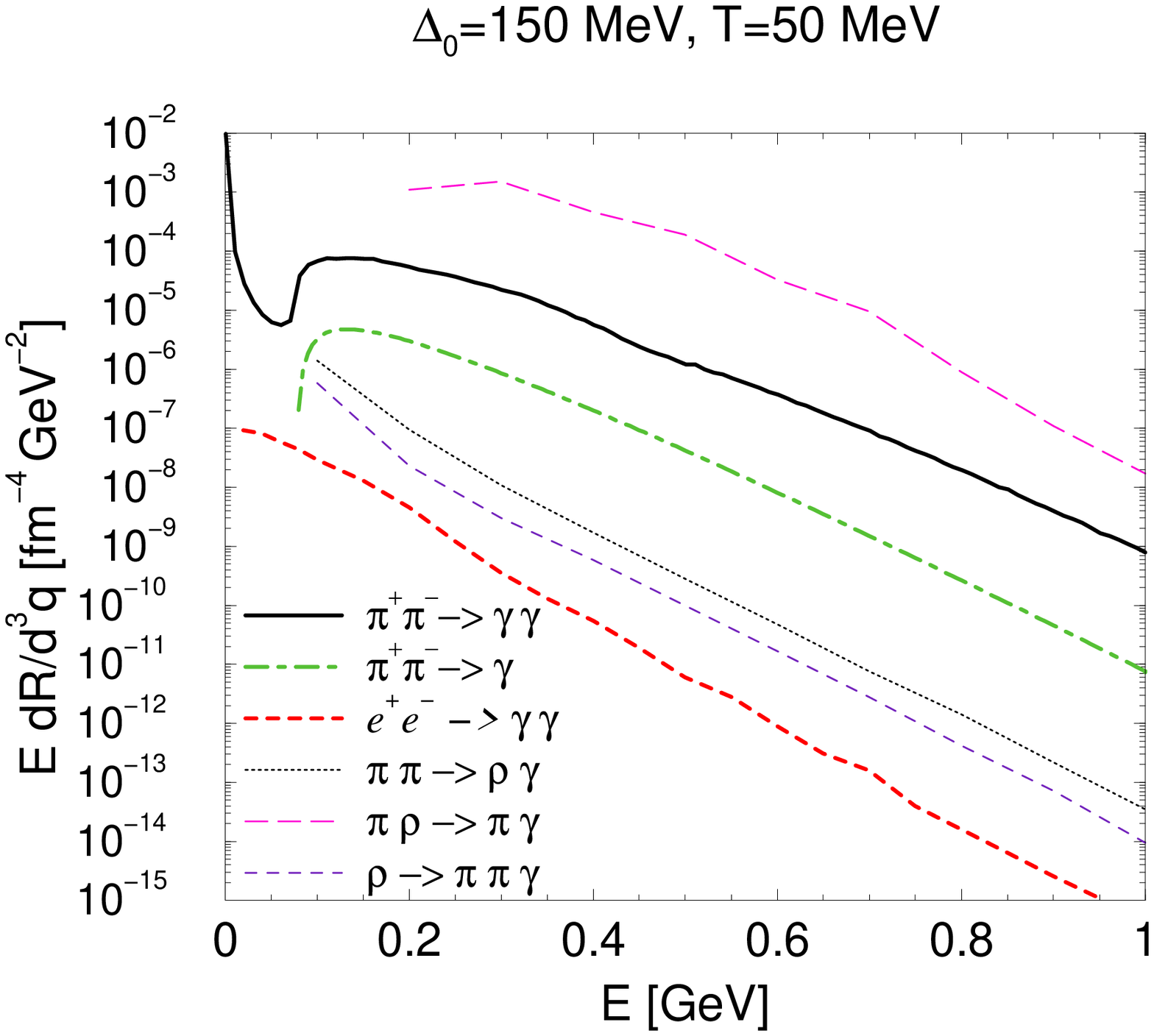,width=0.49\textwidth}
\psfig{file=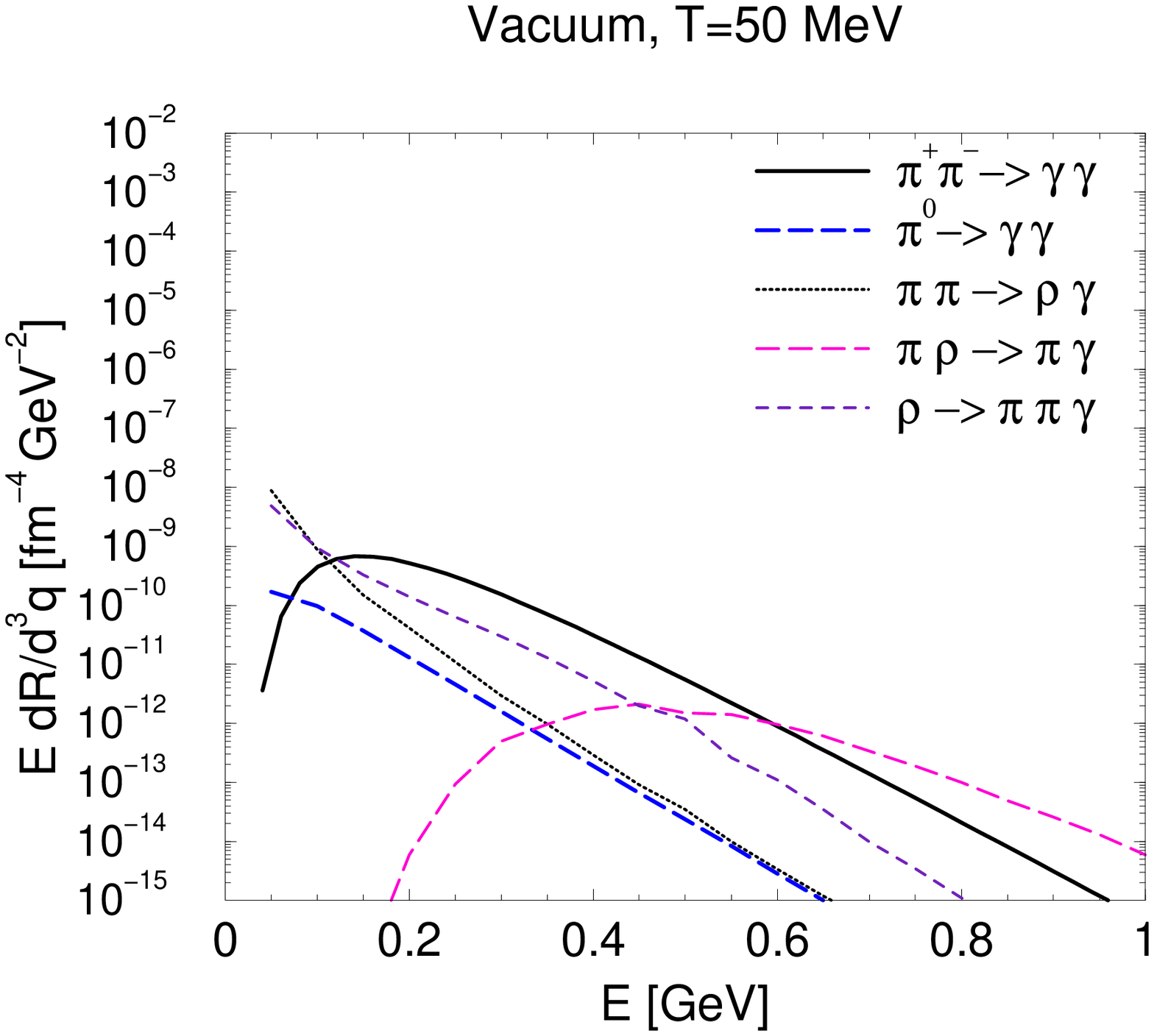,width=0.49\textwidth}
\caption{Photon emission rates in CFL matter from various electromagnetic 
(thick lines) and strong-interaction (thin lines) sources as a function 
of the photon energy $E$. Upper plots:  $\Delta_0 = 50 \mev$ (left) and 
$\Delta_0 = 150 \mev$ (right) at $T = 20 \mev$;  lower plots:  
$\Delta_0 = 150 \mev$ (left) and vacuum phase (right) at $T = 50 \mev$.}
\label{fig:rates}
\end{center}
\end{figure}

The results for the emission rates in CFL matter for different gap 
values and temperatures are summarized in Fig.~\ref{fig:rates}. 
As a check of our numerical calculation, we have ensured agreement
with the results of Refs.~\cite{kap:1991,song:1993} when using the
vacuum values for masses (and velocities) of the pseudoscalar and
vector mesons (pertinent curves for $T=50 \mev$ are displayed in 
the lower right panel of Fig.~\ref{fig:rates}). 
Comparing the lower two panels, one observes that photon emission 
from CFL matter is strongly enhanced, which is essentially due to the
increased phase space, as mentioned above. The particularly large rates
for the processes $\pi^+ \pi^- \to \gamma \, \gamma$ and 
$\pi \rho \to \pi \gamma$ have their origin in the modified pion
dispersion relation to be discussed in Sec.~\ref{sec:lbreaking}. 
The upper two panels indicate that strong processes 
involving rho mesons are further enhanced for smaller gap values, 
due to the reduced rho-meson mass. The emission rate for 
$\pi^0 \to \gamma \, \gamma$ is only shown in the vacuum case.
In CFL matter this rate is kinematically suppressed: it vanishes 
if the photon energy is larger than $m_\pi/\sqrt{1-v_\pi^2}$.
We finally remark that
finite size effects of the effective meson states, \ie, vertex 
form factors, have not been included here (for a recent assessment of 
their suppression effects in photon production at high energies 
in normal hot hadronic matter, cf.~Ref.~\cite{Turbide:2003si}).    

\begin{figure}[!tb]
\begin{center}
\psfig{file=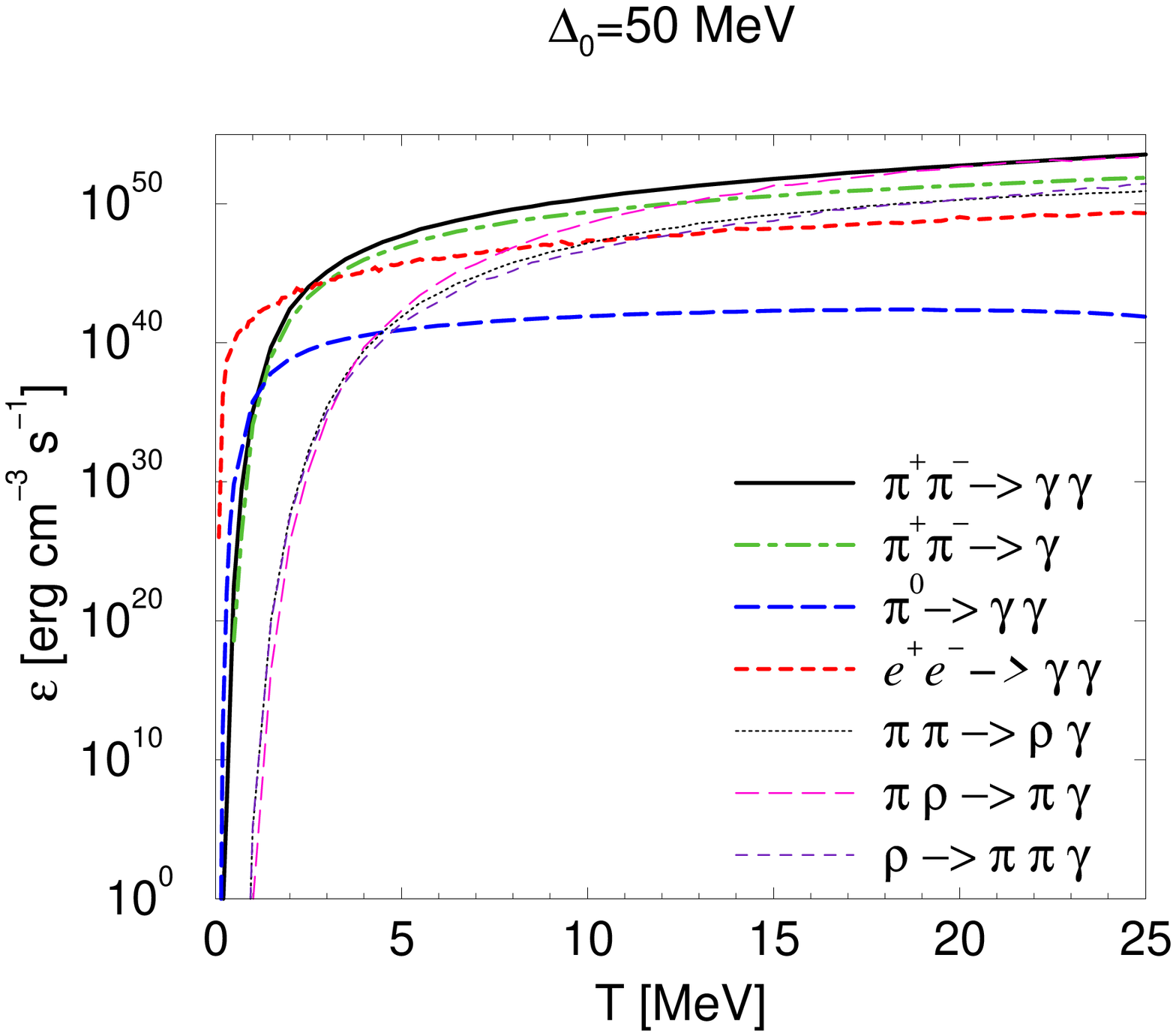,width=0.49\textwidth}
\psfig{file=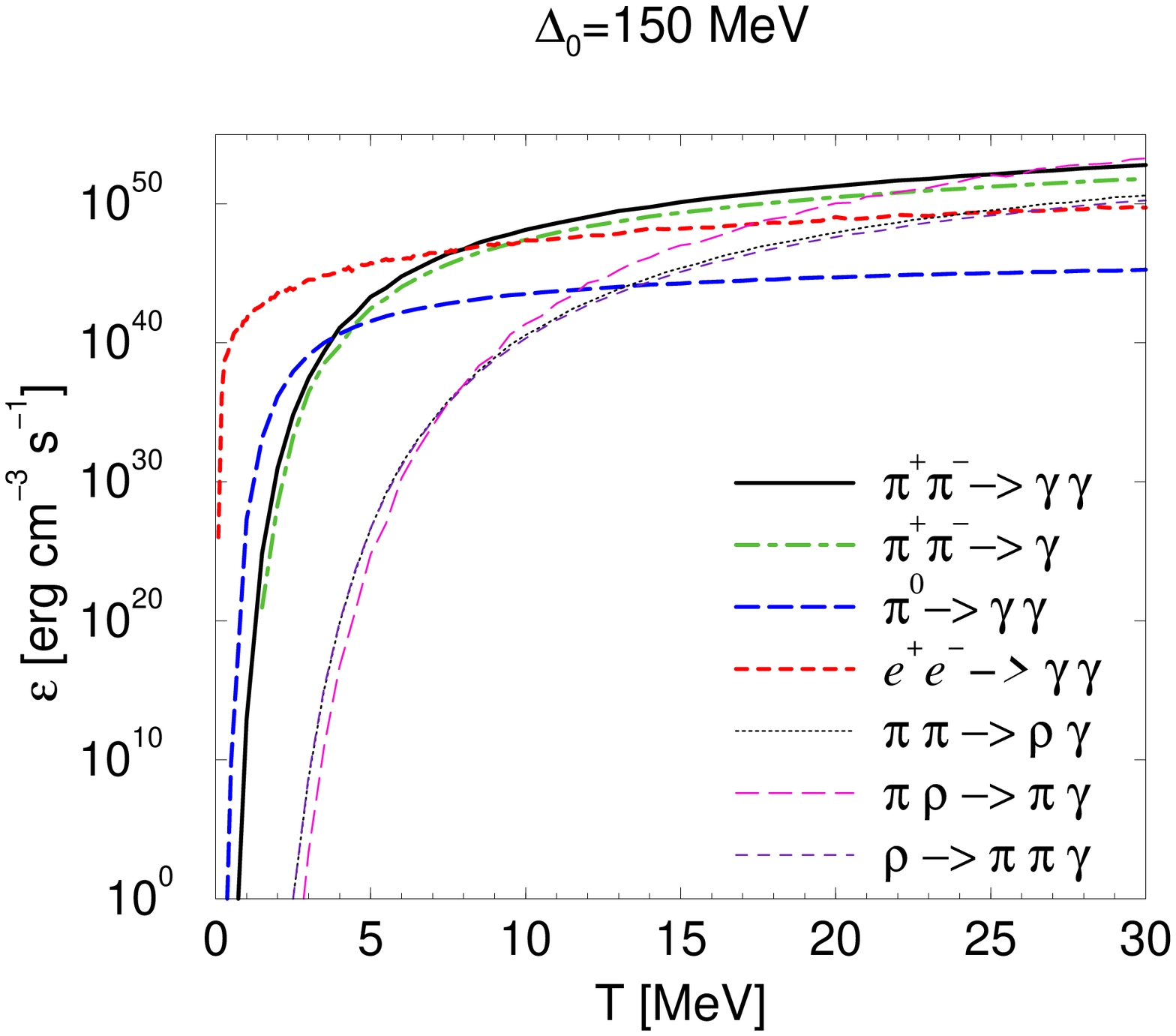,width=0.49\textwidth}
\caption{Photon emissivities in CFL matter from various electromagnetic
(thick lines) and strong-interaction (thin lines) sources as a function
of the temperature for $\Delta_0 = 50 \mev$ (left) and
$\Delta_0 = 150 \mev$ (right).}
\label{fig:emissivity}
\end{center}
\end{figure}
The total photon emissivities, \ie, the emitted photon energies per 
unit volume and time, are now easily obtained by integrating the 
emission rate~(\ref{eq:rate}) (multiplied by the corresponding 
Bose enhancement factor) over the photon 3-momentum, 
\ba
\label{eq:emissivity}
 \epsilon_\gamma = \int d^3 q \, \bigg( E \, \frac{dR}{d^3 q} \bigg) \, 
  [ 1 + f(E,T)] \ .
\ea
Note that due to rotational symmetry the angular integration in the 
variable ${\bf q}$ can be done trivially yielding a factor $4 \, \pi$. 
The photon emissivity for all processes in CFL matter presently
investigated is shown in Fig.~\ref{fig:emissivity}. The process 
$e^+ e^- \to \gamma \, \gamma$ dominates up to temperatures of 
about 3 (7)~\mev\ for $\Delta_0 = 50$ (150)~\mev.
For larger $T$, $\pi^+ \pi^- \to \gamma \, \gamma$ prevails. 
Strong processes involving rho mesons start to compete at 
temperatures of about 20 (25)~\mev\ for $\Delta_0 = 50$ (150)~\mev. 
However, the rho contributions are rather sensitive to the value 
of the rho-mass, subject to appreciable uncertainties due to, \eg,
(i)  sizable corrections to the leading-log estimate,
Eq.~(\ref{eq:mrho}),
(ii) finite width effects (not included at present) from the 
$\rho \leftrightarrow \pi\pi$ coupling, which extend the $\rho$ 
spectral strength down to the two-pion threshold. Therefore, our 
current calculations cannot yet reliably determine at which 
temperatures strong processes outshine the electromagnetic ones.
Finally, let us point out that, due to the (partial)
inversion of the mass hierarchy in the strange and 
nonstrange sector referred to in Sect. 2.1, we expect 
contributions from kaon pairs to be important or even
dominant. These are not included at present, but can 
be treated by a straightforward extension of our calculations
to $SU(3)_F$.

\subsection{Mean free path}

\begin{figure}[t]
\begin{center}
\psfig{file=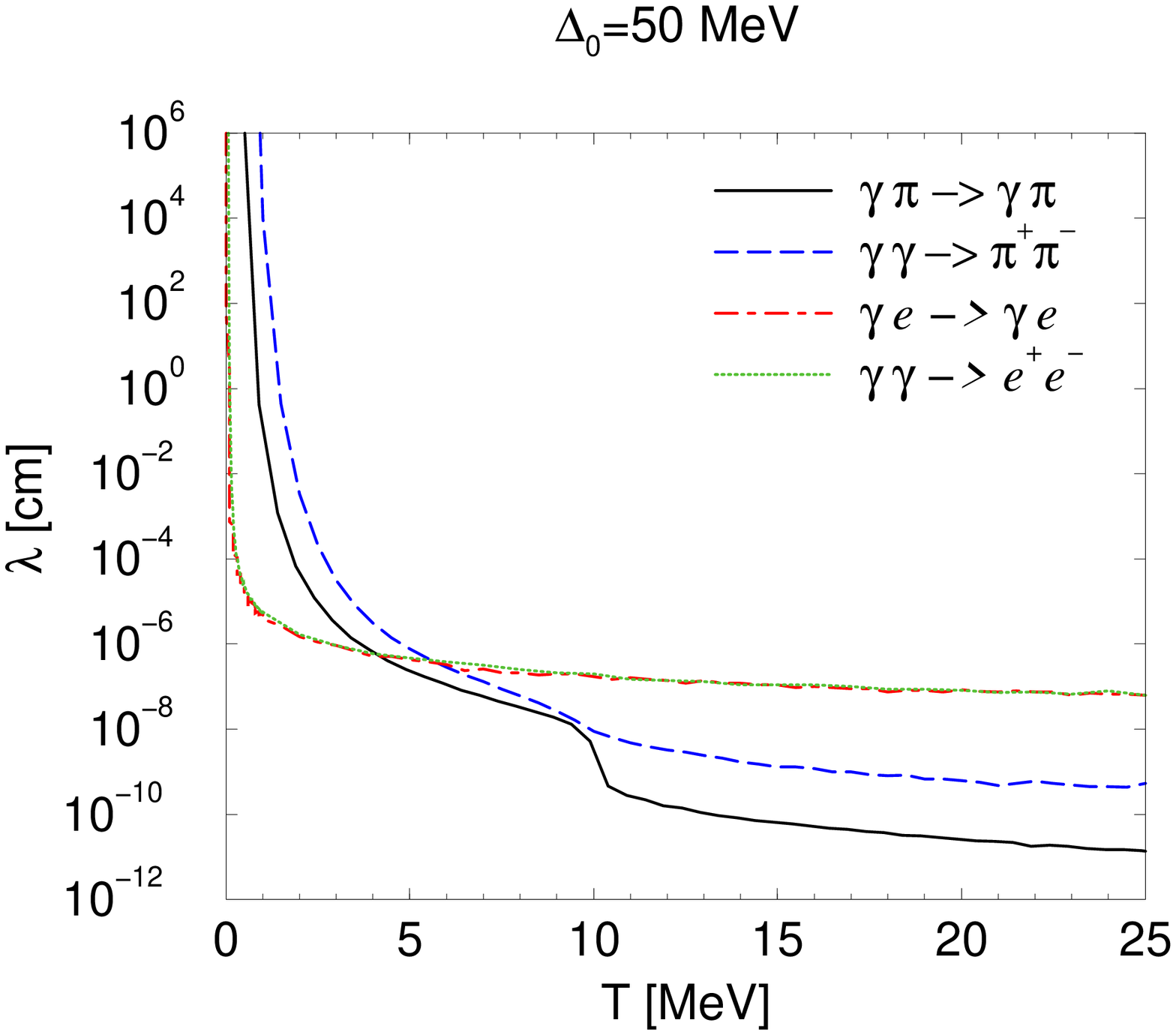,width=0.49\textwidth}
\psfig{file=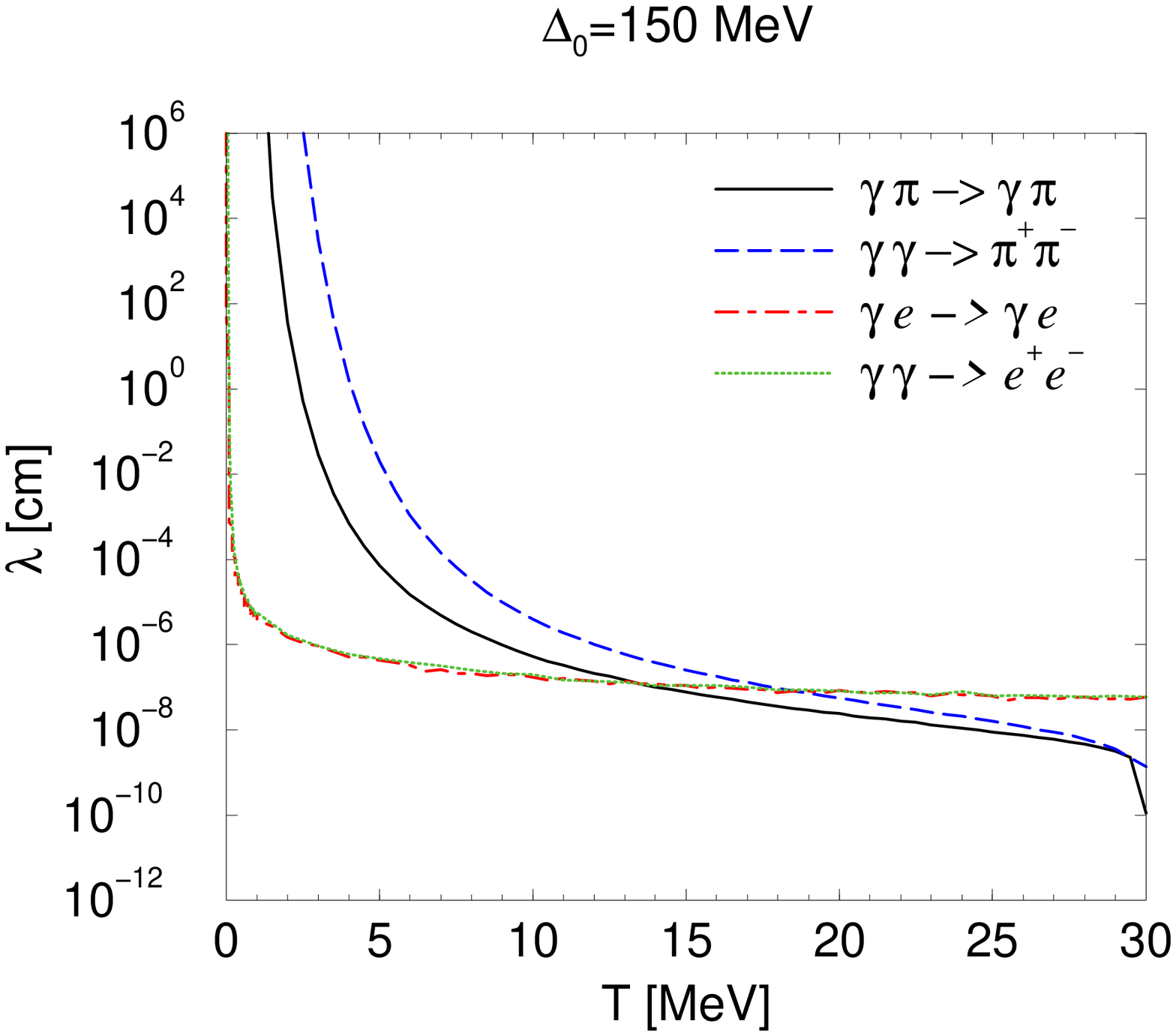,width=0.49\textwidth}
\caption{Photon mean free path from Compton scattering off $\pi^\pm$,
$e^\pm$, and photon annihilation into $e^\pm$ and $\pi^\pm$ pairs,
for $\Delta_0 = 50 \mev$ (left) and $\Delta_0 = 150 \mev$ (right).}
\label{fig:mfp}
\end{center}
\end{figure}

To estimate the emitted photon luminosity of a finite size object
an assessment of the pertinent mean free path is mandatory.
In CFL matter, the latter is governed by Compton scattering off the
light charged excitations, as well as photon annihilation reactions.
In the $SU(2)$ flavor sector, the lightest charged excitations are 
the pions as well as electrons and positrons. Similar to the emissivity 
discussed in the previous Section, at low $T\sim 1~\mev$ one can expect 
processes involving $e^\pm$ to be dominant 
(due to their relatively large number densities), 
whereas towards higher $T$ reactions involving pions 
ought to take over. As mentioned in the preceding subsection, 
another contribution arises from the novel ``decay'' process
$\gamma \to \pi^+ \pi^-$, which owes its existence to the 
in-medium modified pion dispersion relation. 

The inverse \mfp\ of a photon of energy $E$ is given by its total 
cross section on all medium particles per unit volume (schematically 
$\langle \sigma_{\gamma X} \, n_X \rangle $ for each particle
species $X$ of density $n_X$).
For Compton scattering, $\gamma(q)\, X^\pm(p_1) \to
\gamma(q')\, X^\pm(p_2)$, one has
\ba
\frac{1}{\lambda(E)} &=& \frac{1}{2 E}
 \int\frac{d^3 {\bf p}_1}{(2\pi)^3 \, 2 E_1}
 \int\frac{d^3 {\bf p}_2}{(2\pi)^3 \, 2 E_2} 
 \int\frac{d^3 {\bf q}'}{(2\pi)^3 \, 2 E'} \, 
 (2\pi)^4 \, \delta^{(4)}(q + p_1 - q' - p_2) \nn \\
 &&\times \Big| {\cal M}^{\gamma X^\pm \to \gamma X^\pm} \Big|^2
  \, f(E_1,T) \, [ 1\pm f(E_2,T)] \, [ 1+f(E',T)] \,,
\ea
where the upper (lower) sign holds for $X$=$\pi$ ($e$) in connection
with Bose-Einstein (Fermi-Dirac) distributions $f(E_1,T)$ and $f(E_2,T)$. 
Here, the process amplitude ${\cal M}$ is averaged over the initial photon
polarisation and summed over all others. An analogous expression
holds for $\gamma\gamma\to e^+e^-$ annihilation.

In Fig.~\ref{fig:mfp} we display our results for the \mfp\
at a typical thermal energy of the incoming photon, $E=3 \, T$.  
We confirm that for $\Delta_0 = 50$ (150) \mev\ Compton scattering
off $e^\pm$ prevails for temperatures of up to about 5 (15) \mev,
while at larger temperatures, charged pions are the more relevant 
scatterers. Photon ``decay'' into two pions
can only occur if the photon energy $E \ge 2 \, m_\pi/(1-v_\pi^2)$, 
which for $E = 3 \, T$ implies a minimum temperature 
$T_{\rm min} = m_\pi$. Hence, this novel process does not play a 
significant role at low temperatures. As we will elaborate in the 
following Section, it becomes however relevant for $T \ge m_\pi$. 
This induces the rather sharp decrease 
of the solid line in Fig.~\ref{fig:mfp} around $T\simeq 10 \mev$ (left) 
and $T\simeq 30 \mev$ (right), corresponding to the pion mass for 
$\Delta_0 = 50 \mev$ and $\Delta_0 = 150 \mev$, respectively.

\subsection{In-Medium Pion Dispersion Effects}
\label{sec:lbreaking}

\begin{figure}[t]
\begin{center}
\psfig{file=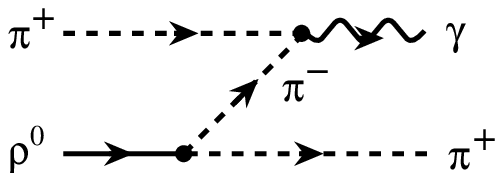,width=0.35\textwidth}
\psfig{file=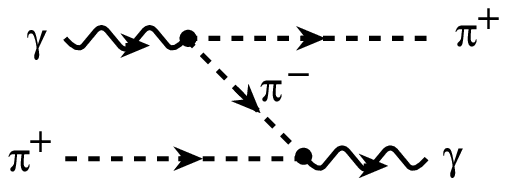,width=0.35\textwidth}
\caption{Z-graph contribution to $\pi^+ \rho^0 \to \pi^+ \gamma$ and
$\gamma \, \pi^+ \to \gamma \, \pi^+$. In time-ordered perturbation
theory these diagram have a cut with three pion states.
Arrows are pointing in the direction of momentum flow.}
\label{fig:z-graph}
\end{center}
\end{figure}

We now return to the specific effects of the in-medium pion velocity
which are at the origin of  the novel emission and absorption 
processes, $\pi^+ \pi^- \to \gamma$ and 
$\gamma \to \pi^+ \pi^-$, respectively, already mentioned above. 
To illustrate their appearance in our calculations, let us for 
definiteness consider the reaction $\pi^+ \rho^0 \to \pi^+ \gamma$.  
This process has a contribution from a $t$-channel pion-exchange diagram.
According to Eq.~(\ref{eq:pion-prop}), the pertinent propagator, 
$D_\pi$, is modified by a factor $v_\pi^2$ in the denominator, so 
that it takes the explicit form 
\ba
 D_\pi^{-1}(p_0,{\bf p}) = p_0^2 - v_\pi^2 \, {\bf p}^2 - m_\pi^2 
 = E^2 \, (1-v_\pi^2) - 2 \, E \, (E_1 - v_\pi \, \sqrt{E_1^2-m_\pi^2} \, 
 \cos\Theta_1) \,.
\label{eq:denom}
\ea
As before, $E$ is the final-state photon energy, $E_1$ the 
initial-state pion energy, $\Theta_1$ the polar angle 
between their three momenta (the $z$-axis is chosen
along the direction of the photon momentum), and
$t=p^2=(p_1-q)^2$ is the four-momentum of the exchanged 
{\em positively} charged pion.  In vacuum ($v_\pi=1$), 
$D_\pi^{-1}$ is always negative, \ie, in the spacelike regime. 
In the medium, however, the presence of the term 
$E^2 \, (1-v_\pi^2)$ implies that $D_\pi^{-1}$ can become timelike 
and therefore develop singularities in the physical region.  
Evidently, timelike values can only occur for sufficiently large 
photon energies. If, \eg, the incoming pion is at rest, 
$D_\pi^{-1}=0$ requires a minimal photon energy of 
$E_{\rm min} = 2 \, m_\pi / (1-v_\pi^2)$.
Furthermore, in this situation the intermediate 
pion carries negative energy, readily interpreted 
as being its antiparticle with positive energy and reversed three 
momentum, that is, an {\em incoming}, {\em on-mass-shell}, $\pi^-$. 
In the language of time-ordered perturbation theory this process
represents a {\it Z-graph} contribution, depicted in the left panel of 
Fig.~\ref{fig:z-graph}, where the incoming $\pi^+$ annihilates with an 
intermediate $\pi^-$ to produce the outgoing photon. The final 
scattering off the incoming $\rho^0$ to bring the $\pi^+$ on-shell is 
no longer required,  contrary to the vacuum case. 
In vacuum, the virtual $\pi^+$ only goes on-mass shell in the limit
of zero photon energy, in which case the phase space vanishes,
and the emissivity receives no contribution.
Completely analogous arguments apply for other charge combinations of 
the process $\pi \, \rho \to \pi \, \gamma$. 

One can verify the above features more explicitly by directly
analyzing the process $\pi(p_1) + \pi(p_2) \to \gamma(q)$.
Exploiting both energy and momentum conservation, $E_1 + E_2 = E$ and 
${\bf p}_1 +{\bf p}_2 = {\bf q}$, together with the on-mass shell
dispersion relations for all three particles involved,
one finds for $\cos\Theta_1=\pm 1$:
\be
E^\pm=3 \, E_1 \pm |{\bf p}_1| \ ,
\ee 
where we have used that $v_\pi^2 = 1/3$.
This, in particular, implies that for an outgoing photon its (positive)
energy $E$ is always larger than $E_1$ and therefore $E_2 > 0$, that is,
the Bremsstrahlungs-process $\pi(p_1) \to \pi(p_2) \, \gamma(q)$ is not
possible.\footnote{Equivalently, one can start from energy and momentum
conservation for the Bremsstrahlungs-diagram to find that $E_2$ must
be negative, which again means that only the annihilation process
can occur.}

A similar discussion of the in-medium pion dispersion effects pertains 
to processes that contribute to the mean free path. 
As we have already pointed out in the previous Section, 
at a temperature corresponding to the pion mass the 
pion-induced mean free path exhibits a rather 
sharp decrease. In our calculation this is associated with the 
$u$-channel diagram in $\gamma \, \pi$ Compton scattering, 
which exhibits a behaviour analogous to the $t$-channel 
diagram discussed above, \ie, the intermediate pion becomes
on-shell with negative energy.   
The interpretation via the corresponding Z-graph is that the incoming 
photon spontaneously decays into two on-shell charged pions, 
cf.~the right panel of Fig.~\ref{fig:z-graph}.   

For the numerical evaluation of the diagrams involving on-shell 
intermediate pions we include in the (singular) pion-exchange propagators 
a finite width that can be attributed, \eg, to interactions with 
surrounding medium pions. A schematic estimate of the pion width is given  
in Appendix~\ref{app_pionwidth}. For the present purpose it is sufficient 
to approximate the width by a constant; in the kinematical range 
of interest the evaluation of $\rho\pi\pi$ interactions yields 
as an upper estimate a value of 
$\Gamma_\pi \simeq 0.1 \mev$ for $\Delta_0 = 50 \mev$ and 
$\Gamma_\pi \simeq 0.01 \mev$ for $\Delta_0 = 150 \mev$.

\subsection{Comparison with Neutrino Properties}
\label{sec:neutrinos}

At this point, a comparison with the results of Ref.~\cite{red:2002}
is in order, where an analogous investigation of neutrino emissivities
and mean free paths was presented. On the one hand, we find that the 
photon emissivity exceeds the neutrino emissivity by more than ten orders 
of magnitude over the entire temperature region. On the other hand,
as to be expected for weakly interacting neutrinos,  their mean free 
path is larger than that of photons by several orders of magnitude.  

\begin{figure}[t]
\begin{center}
\psfig{file=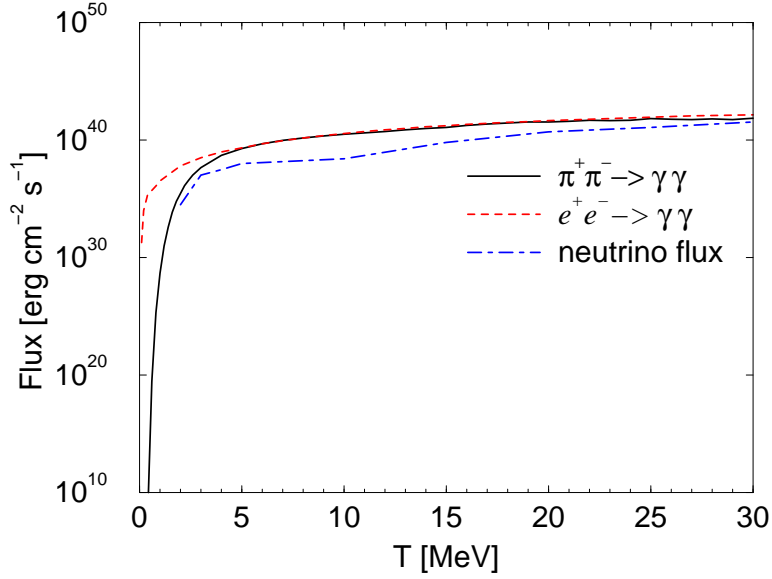,width=0.6\textwidth}
\caption{Comparison of photon and neutrino fluxes, with the parameter choice
of Ref.~\cite{red:2002}, $\Delta_0 = 100 \mev$ and $\mu_q = 400 \mev$.}
\label{fig:flux}
\end{center}
\end{figure}

To confront the efficiencies of photon and neutrino emission
in the radiation of energy from a hypothetical CFL star, we have 
evaluated the corresponding fluxes as follows. In order to gain a 
first estimate we ignore the possibility of a nuclear crust which
might shield the CFL phase.
For large opacities, as implied by the results found above, 
the photon flux can be obtained from a convolution of 
the emissivity and mean free path, integrated over the photon energies. 
The neutrino flux, for simplicity, has been estimated by multiplying 
the emissivities and mean free paths as given in the first entry of 
Ref.~\cite{red:2002}, which amounts to the assumption that the neutrinos 
are thermalized when leaving the star.\footnote{This is a good 
approximation for temperatures above a few MeV~\cite{red:2002}.}
To facilitate a direct comparison with the findings of Ref.~\cite{red:2002},
we have employed their values for the gap and the quark chemical potential,
$\Delta_0 = 100 \mev$ and $\mu_q = 400 \mev$, respectively. 
The results are displayed in Fig.~\ref{fig:flux}, indicating that, 
despite of the extremely small mean free path of the photons, their 
emitted energy flux exceeds that of neutrinos by one to three orders 
of magnitude over a wide temperature range.\footnote{Here we are 
implicitly assuming that photons and neutrinos are emitted at the 
same temperature. This does not hold, however, once a significant 
temperature gradient has established at the surface. In that case, 
neutrinos will be predominantly emitted from hotter regions in the 
interior, due to their larger mean free path.} 

Note that the photon flux approximately corresponds to that of a 
black-body emitter. This complies with the general expectation that, 
for very small photon mean free path, emission and absorption processes 
are close to equilibrium and reflect the surface temperature of the star.  

\section{Early Thermal Evolution}
\label{sec:evo}

In this Section, we present a rough estimate of the early thermal 
evolution of a hypothetical CFL star. 
The thermal properties of CFL matter are expected to be dominated by the 
massless Goldstone boson $\phi$, which results from the
breaking of the baryon number symmetry $U(1)_B$. The corresponding specific
heat has been given in Ref.~\cite{jai:2002}, 
\ba
 c_{v,\phi} = \frac{2 \sqrt{3} \, \pi^2}{5} \, T^3
 = 7.8 \times 10^{16} \, T^3_{\mev} \erg \cm^{-3} \K^{-1} \,.
\ea
The thermal conductivity in a CFL phase has been estimated  
in Ref.~\cite{sho:2002a}, where the contribution from the massless 
Goldstone boson was found to be
\ba
 \kappa_\phi=1.2 \times 10^{27} \, T^3_{\mev} \, \lambda_{\rm cm} \,
 \erg \cm^{-1} \s^{-1} \K^{-1} \,,
\ea
with $\lambda_{\rm cm}$  being the mean free path of the \gb\ in cm.
The dominant process for the \mfp\ is the
decay of \gb\ into pairs of quark quasiparticles. Assuming thermalized
Goldstone modes, the mean free path reads~\cite{sho:2002a}
\ba
 \lambda_{\phi \to qq}(T) \simeq \frac{4 (21-8 \ln 2)}{15\sqrt2\,\pi\, T}
 \, \exp\Bigg( \sqrt{\frac32} \frac{\Delta}{T} \Bigg) \,.
\ea
Note that $\lambda$ becomes large, \ie\ greater than about 1~km, only
for temperatures below 5 MeV.

Since, at large $T$, the thermal conductivity is extremely small, we may
estimate the cooling time of the CFL surface in the very early stage by
neglecting the heat transport in the interior of the star. We thus use
the thermal energy, $U_{\rm th,shell}$,  stored in a shell with a 
thickness given by the photon mean free path, 
$\Delta R \simeq \lambda_\gamma$, to write the luminosity as
\ba
 \frac{d U_{\rm th,shell}}{d t} = - L_\gamma \,.
\label{eq:econs}
\ea
$U_{\rm th,shell}$ is readily obtained by integrating the heat capacity
over temperature and multiplying with the volume of the surface shell,
\ba
 U_{\rm th,shell} =
 4 \pi R^2 \, \lambda_\gamma \int_0^T c_{v,\phi}(T') \, dT' \,.
\label{eq:e_thermal}
\ea
Using the black-body luminosity, $L_\gamma = 4 \pi R^2 \, \sigma \, T^4$,
together with Eq.~(\ref{eq:e_thermal}) in expression~(\ref{eq:econs}), 
and setting the radius of the star to $R = 10 \km$, we find that the 
cooling time $\Delta t$ of the shell is of the order
\ba
 \Delta t \sim 10^{-16} \s \, \ln\frac{T_i}{T_f} \,,
\ea
with $T_i$ being the initial and $T_f$ the final temperature. This cooling
time scale is extremely short, and is to be compared with the
diffusion time of the massless \gb, corresponding to the thickness
$\Delta R$ of the shell. The latter time scale might be assessed by 
the simple relation (see, \eg, Ref.~\cite{sha:1983})
\ba
 t_D = \bigg( \frac{\lambda_\gamma}{\lambda_\phi} \bigg)^2 \, t_{\rm int} \,,
\ea
where $t_{\rm int} \simeq 10^{-20} \s$ is the typical interaction time of
the massless \gb\ (estimated from their decay $\phi \to q q$) at
$T \sim 30 \mev$. With $\lambda_\phi(T = 30 \mev) \sim 10^{-10} \cm$ we
find $t_D \sim 10^{-14} \s$, which is much longer than the cooling time
$\Delta t$. The large difference in these time scales justifies the 
neglect of the heat transport in the early stages of the cooling.

\section{Summary and Conclusions}
\label{sec:sum}

The main objective of this work has been an order of magnitude assessment
of photon emission rates and mean free paths in a CFL phase of high-density
quark matter. Our approach 
is based on a low-energy effective theory that describes the prevailing
degrees of freedom at large densities and moderate temperatures, 
$T \sim \morder{10 \mev}$. These are generalized \gb\ that appear due to 
breaking of chiral and baryon number symmetries. We have followed the 
hidden local symmetry approach to introduce vector-meson excitations 
as well as the electromagnetic field in our analysis. 
The assumption of the KSRF relation in medium enabled us to relate
the vector-meson self-interaction strength $\widetilde{g}$ and two-pion
coupling $\grho$ to the characteristic scales of the CFL phase,
\ie, the gap $\Delta$ and the quark chemical potential $\mu_q$, 
according to $\grho=\wt{g}\simeq \sqrt2\, \Delta/f_\pi$, where 
$f_\pi \simeq 0.21 \, \mu_q$. In the presence of in-medium modified
pion dispersion relations, electromagnetic gauge invariance could
be maintained by requiring both the $\rho\,$-$\,\gamma$ vertex as 
well as the rho propagator (with mass $m_\rho\simeq 2\Delta$) 
in its vacuum form (which emerged as a consequence of neglecting
the distinction between temporal and spatial components of the
rho mass). 

The resulting photon production and scattering amplitudes to lowest 
order have been convoluted over available phase space in CFL matter 
using standard kinetic theory expressions. 
Our results suggest that photon emissivities might become
very large, up to $\morder{10^{53} \erg \cm^{-3} \s^{-1}}$
at the highest temperatures expected to occur during (or immediately 
following) a supernova explosion, $T\simeq 30 \mev$~\cite{sha:1983}. 
Over a large range of CSC gap values, $\Delta_0=$ 50 -- 150~\mev, we find
that at temperatures below $\sim$5~\mev\ the dominant process is 
$e^+ e^- \to \gamma \, \gamma$, whereas for larger $T$ electromagnetic 
pion annihilation, $\pi^+ \pi^- \to \gamma \, \gamma$, is most 
important. Processes invol\-ving strongly interacting rho mesons could 
also play a major role. The temperature where the strong processes 
begin to outshine electromagnetic ones could not be accurately determined 
(although it might be well below 30~MeV), mostly due to 
uncertainties in the vector-meson mass spectrum.   
An interesting feature of the CFL phase is the appearance of the
novel processes $\pi^+ \pi^- \leftrightarrow \gamma$, which are 
triggered by the in-medium pion dispersion relation due to the
breaking of Lorentz invariance.

We furthermore have found that the photon mean free path is extremely 
small over a wide range of temperatures. In particular, it is far below 
the typical radius of a compact star as long as the temperature is 
above about a few tens of keV.
With a gap value of 50 (150)~\mev, the processes providing the largest 
opacity for $T$ up to 5 (15)~\mev\ are Compton scattering off thermal 
electrons and positrons, $\gamma \, e^\pm \to \gamma \, e^\pm$, as well 
as photon annihilation, $\gamma \, \gamma \to e^+ \, e^-$, in 
agreement with previous calculations in a low-temperature 
approximation~\cite{sho:2002b}. 
At larger $T$, Compton scattering off pions and photon annihilation into 
pion pairs are dominant.

The comparison with analogous calculations of neutrino properties
in CFL matter~\cite{red:2002} revealed that the photon emissivity 
is much larger than the neutrino one, whereas the mean free path
is substantially shorter. A schematic estimate of the photon flux from 
a hypothetical CFL star with uniform temperature profile indicates that 
the photon flux could even exceed the volume-driven neutrino flux 
(temperature gradients will favor, however, the latter).  

We have also presented an order of magnitude analysis of the cooling time
of the outermost surface layers of a (hypothetical) CFL star,
in the very early stages of its evolution. Since, at the highest 
temperatures considered here, 
the thermal conductivity of CFL matter is very small, one may
neglect heat transport in the interior of the star. The thus found cooling
time is of the order $10^{-16} \s$. This is to be compared with the diffusion
time of a massless Goldstone boson, $t_D \sim 10^{-14} \s$, corresponding
to the distance given by the photon mean free path at large temperature. 
These widely different time scales suggests that heat transport does not 
play a significant role in the very early cooling era.

Let us finally comment on a few aspects of our work that have not 
been adequately treated and should be addressed in future analyses: 

(i) In our calculation of the emission rates and mean free paths we 
have restricted ourselves to the $SU(2)$-flavor case, \ie,  did not include 
contributions from strangeness-carrying mesons in the pseudoscalar and 
vector octet. The analysis of 
Ref.~\cite{sch:2001} showed that, while the $K^-$ is slightly heavier 
than the pion, the $K^+$ is considerably lighter. Since the sum of the 
kaon masses is less than the sum of the pion masses, one can expect 
that the emission rates from $K^+ K^-$ annihilation into photons will 
exceed the emission rates resulting from $\pi^+ \pi^-$ annihilation. 
Likewise, the kaon induced mean free path is expected to be smaller 
than the one induced by pions. 
Following our arguments of Sect.~\ref{sec:neutrinos} we can assume, 
however, that the overall result for the flux will remain unchanged,
being limited by black-body emission.
At moderate densities, where the pion decay constant is of the order
of the gap, one can also expect solitonic excitations in the baryon
sector~\cite{Jackson:2003dk} to become relevant. 

(ii) Our discussion of the cooling behavior of a compact star featuring 
CFL matter was limited to the very early stages of the thermal 
evolution, and we have furthermore made the simplifying assumption that 
the star accommodates a CFL phase which extends all the way up to the 
surface.  We will postpone the investigation of the long-term cooling, 
including heat transport and more realistic scenarios in which the CFL 
phase is covered with a nuclear crust, to a future 
publication~\cite{ouy:2003}. 

\vskip\baselineskip

{\bf Acknowledgements.}
We would like to thank Emma Olsson, Chris Pethick, Robert Pi\-sarski, 
Igor Shovkovy, Kim Splittorff and Dmitri Voskresensky for interesting 
discussions. We are especially indebted to Francesco Sannino for his 
continuous encouragement throughout the work.
C.V. wishes to thank Deirdre Black and Sanjay Reddy for many enlightening
conversations, and the Theory Group at Jefferson Lab for their kind
hospitality during his visit, where parts of this work have been done.
He also thanks Igor Musatov for providing his program FeynmanGraph.
This work is partially funded by the European Commission IHP program under 
contract HPRN-CT-2000-00130. The research of R.~O. is supported by grants 
from the Natural Science and Engineering Council of Canada (NSERC). 

\appendix
\section{Appendix}
\subsection{In medium Feynman rules}
\label{app_feynman}

The Feynman rules follow from the effective Lagrangian discussed in 
Sect.~\ref{sec:effL} in a straight-forward fashion. From the kinetic 
term in Eq.~(\ref{eq:chilgn}) we obtain the pion propagator,   
\ba
 i\, D_\pi= \frac{i}{p_0^2 - v_\pi^2\, {\bf p}^2 -m_\pi^2 + i\, \epsilon} 
\ .
\label{eq:pion-prop}
\ea

The vertex factors derive from the interaction part~(\ref{eq:intlgn}).
For the $\pi^+(p) \rho^0 \to \pi^+(p')$ vertex we find
\ba
 i \, \grho \, ( \wt{p} + \wt{p}\,')_\mu \ ,
\ea
where $\wt{p}= (p_0, v_\pi^2 \, {\bf p})$. 
The $\pi^+ \pi^- \rho^0 \gamma$ contact vertex reads
\ba
 2 \, i \, e \, \grho \, \wt{g}_{\mu\nu} \ ,
\ea
and for $\pi^\pm \pi^0 \rho^\pm \gamma$ we have
\ba
 - i \, e \, \grho \, \wt{g}_{\mu\nu} \ ,
\ea
with $\wt{g}_{\mu\nu}= {\rm diag}(1,-v_\pi^2,-v_\pi^2,-v_\pi^2)$.
The $\pi^+ \pi^- \gamma \gamma$ contact term is
\ba
 - 2\, i \, e^2 \, \wt{g}_{\mu\nu} \ .
\ea
For the $\rho^0\,$-$\,\gamma$ coupling we obtain
\be
 - i \, \frac{e\, m_V^2}{\wt{g}} \, g_{\mu\nu} \ .
\ee
Finally, the triple-rho vertex is given by
\ba
 i\, \grho\, [ g_{\mu\nu} \, (p-p')_\lambda + g_{\nu\lambda} \, (p'-q)_\mu +
  g_{\lambda\mu} \, (q-p)_\nu ] \ ,
\ea
where all momenta are flowing {\it into} the vertex.

\subsection{Process amplitudes}
\label{app_amplitude}

The process amplitudes for
$\pi^+(p_1) \, \pi^-(p_2) \to \gamma(q_1) \, \gamma(q_2)$ read
\ba
 {\cal M}_{(1)}^{\mu\nu} &=& e^2 \, 
 \frac{(2 \, \wt{p}_1-\wt{q}_1)^\mu \, (2\, \wt{p}_2 - \wt{q}_2)^\nu}{
 (E_1-q_1)^2 - v_\pi^2 \, ({\bf p}_1 - {\bf q}_1)^2 - m_\pi^2 } \ , \\
 {\cal M}_{(2)}^{\mu\nu} &=& e^2 \, 
 \frac{(2 \,\wt{p}_2 - \wt{q}_1)^\mu \, (2 \, \wt{p}_1 - \wt{q}_2)^\nu}{
 (E_2-q_1)^2 - v_\pi^2 \, ({\bf p}_2 - {\bf q}_1)^2 - m_\pi^2 } \ , \\
 {\cal M}_{(3)}^{\mu\nu} &=& 2\, e^2 \, \wt{g}^{\mu\nu} \  .
\ea
By using crossing relations we immediately obtain the amplitudes for
Compton scattering, $\gamma \, \pi^\pm \to \gamma \, \pi^\pm$, and 
$\gamma\,\gamma\to\pi^+\,\pi^-$, relevant for the calculation of the 
photon mean free path.

The amplitudes for $\pi^+(p_1) \, \pi^-(p_2) \to \rho^0(p_3) \, \gamma(q)$ are
\ba
 {\cal M}_{(1)}^{\mu\nu} &=& e \, \grho \,
 \frac{ (2 \, \wt{p}_1 - \wt{q})^\mu \, (2 \, \wt{p}_2 - \wt{p}_3)^\nu}
 {(E_1-q)^2 - v_\pi^2 \, ({\bf p}_1 - {\bf q})^2 - m_\pi^2 } \ , \\
 {\cal M}_{(2)}^{\mu\nu} &=& e \, \grho \,
 \frac{ (2 \, \wt{p}_2 - \wt{q})^\mu \, (2 \, \wt{p}_1 - \wt{p}_3)^\nu}
 { (E_2-q)^2 - v_\pi^2 \, ({\bf p}_2 - {\bf q})^2 - m_\pi^2 } \ , \\
 {\cal M}_{(3)}^{\mu\nu} &=& 2 \, e \, \grho \; \wt{g}^{\mu\nu} \ ,
\ea
where the indices $\mu$ and $\nu$ refer to the photon and rho meson 
polarisation, respectively.
The amplitudes for $\pi^\pm \rho^0 \to \pi^\pm \gamma$ and
$\rho^0 \to \pi^+ \pi^- \gamma$ directly follow from the above
expressions through crossing.

For the processes $\pi^\pm(p_1)\, \pi^0(p_2) \to \rho^\pm(p_3)\, \gamma(q)$ 
we find
\ba
{\cal M}_{(1)}^{\mu\nu} &=&  - e \, \grho \,
 \frac{ (2 \, \wt{p}_1 - \wt{q})^\mu \, (2\, \wt{p}_2 - \wt{p}_3)^\nu}
 {(E_1-q)^2 - v_\pi^2 \, ({\bf p}_1 - {\bf q}^2) - m_\pi^2 } \ , \\
{\cal M}_{(2)}^{\mu\nu} &=& e \, \grho \, 
 \frac{ (2 \, p_3 + q)^\mu \, (\wt{p}_1 - \wt{p}_2)^\nu 
      - (\wt{p}_1 - \wt{p}_2)^\mu \, (2 \, q + p_3)^\nu
      + g^{\mu\nu} \, (\wt{p}_1 - \wt{p}_2) \cdot (q - p_3) }
 { (p_3 + q)^2 - m_V^2 } \ , \nn \\ \\
{\cal M}_{(3)}^{\mu\nu} &=& - e \, \grho \; \wt{g}^{\mu\nu} \ .
\ea
The amplitudes for $\pi^\pm \rho^\mp \to \pi^0 \gamma$, 
$ \pi^0 \rho^\pm \to \pi^\pm \gamma$ and
$\rho^\pm \to \pi^\pm \pi^0 \gamma$ can again be obtained by crossing.

\subsection{Pion width in a dense medium}
\label{app_pionwidth}

The width of the pion, generated by its interaction with surrounding 
medium pions, is defined by ${\rm Im} \, \Sigma_\pi = - m_\pi  \Gamma_\pi$. 
The pion self-energy can be obtained from the standard expression
(see, \eg, Ref.~\cite{Rapp:1995fv}) 
\ba
 \Sigma_\pi(k_0,{\bf k}) = \int \frac{d^3 p}{(2 \pi)^3 \, 2 p_0} \, 
 f^\pi(p_0,T) \, {\cal M}^{\pi\pi \to \pi\pi} \ ,
\label{eq:sigpi1}
\ea
where $f^\pi$ is the Bose-Einstein distribution and 
${\cal M}^{\pi\pi \to \pi\pi}$
the forward-scattering amplitude of pions that scatter via ($s$-channel) 
rho-meson exchange. Upon evaluating ${\cal M}$ in the center-of-mass (cms) 
frame, Eq.~(\ref{eq:sigpi1}) takes the form
\ba 
 \Sigma_\pi(k_0,{\bf k}) = - \frac{2\, \grho^2}{(2 \pi)^2} \, 
 \int \frac{|{\bf p}| \, d \, |{\bf p}|}{p_0} \, f^\pi(p_0,T) \, 
 \int_{s_{\rm min}}^{s_{\rm max}} \, \frac{ds}{|{\bf k}|} 
 \frac{s/4 - m_\pi^2}{s-m_\rho^2+ i \, {\rm Im} \, \Sigma_\rho} \ .
\label{eq:sigpi2}
\ea 
The integration limits for the integral over the squared cms energy $s$ 
are given  by $s_{\rm min, max} 
= 2\, m_\pi^2 + 2\, k_0 \, p_0 \pm 2\, |{\bf k}| \, |{\bf p}|$, 
and 
\be 
 {\rm Im} \, \Sigma_\rho = m_\rho \Gamma_\rho 
 = \frac{\grho^2}{6 \pi} \frac{q^3}{\sqrt{s}} \ , \qquad 
   q^2=\frac{s}{4} -m_\pi^2 \ .
\ee
For simplicity, we here neglect effects of the in-medium pion 
dispersion relation.



\begin{thebibliography}{99}

\bibitem{col:1974}
J.C.~Collins and M.J.~Perry,
Phys.\ Rev.\ Lett.\  {\bf 34}, 1353 (1975).

\bibitem{bar:1977}
B.C.~Barrois,
Nucl.\ Phys.\ B {\bf 129}, 390 (1977); \\
S.C.~Frautschi, in Proceedings of the Workshop on {\em Hadronic
Matter at Extreme Energy Density}, Erice, Italy, Oct 13-21, 1978,
edited by N. Cabbibo and L. Sertorio
(Plenum, New York, 1978), p. 18.

\bibitem{alf:1997}
M.G.~Alford, K.~Rajagopal and F.~Wilczek,
Phys.\ Lett.\ B {\bf 422}, 247 (1998) [hep-ph/9711395].

\bibitem{rapp:1997}
R.~Rapp, T.~Sch\"afer, E.V.~Shuryak and M.~Velkovsky,
Phys.\ Rev.\ Lett.\  {\bf 81}, 53 (1998) [hep-ph/9711396].

\bibitem{pis:1998}
R.D.~Pisarski and D.H.~Rischke,
Phys.\ Rev.\ Lett.\  {\bf 83}, 37 (1999) [nucl-th/9811104].

\bibitem{son:1998}
D.T.~Son,
Phys.\ Rev.\ D {\bf 59}, 094019 (1999) [hep-ph/9812287].

\bibitem{alf:1998}
M.G.~Alford, K.~Rajagopal and F.~Wilczek,
Nucl.\ Phys.\ B {\bf 537}, 443 (1999) [hep-ph/9804403].

\bibitem{sch:2003}
T.~Sch\"afer,
hep-ph/0304281.

\bibitem{jai:2002}
P.~Jaikumar, M.~Prakash and T.~Sch\"afer,
Phys.\ Rev.\ D {\bf 66}, 063003 (2002) [astro-ph/0203088].

\bibitem{red:2002}
S.~Reddy, M.~Sadzikowski and M.~Tachibana,
Nucl.\ Phys.\ A {\bf 714}, 337 (2003) [nucl-th/0203011]; 
Phys.\ Rev.\ D {\bf 68}, 053010 (2003) [nucl-th/0306015].

\bibitem{jai:2001}
P.~Jaikumar, R.~Rapp and I.~Zahed,
Phys.\ Rev.\ C {\bf 65}, 055205 (2002) [hep-ph/0112308].

\bibitem{sha:1983} S.~L.~Shapiro and
 S.A.~Teukolsky, Black Holes, White Dwarfs, and Neutron Stars:
the physics of compact objects (Wiley-Interscience, 1983).

\bibitem{Turbide:2003si}
S.~Turbide, R.~Rapp and C.~Gale,
hep-ph/0308085.

\bibitem{hong:1999}
D.K.~Hong, M.~Rho and I.~Zahed,
Phys.\ Lett.\ B {\bf 468}, 261 (1999) [hep-ph/9906551].

\bibitem{cas:1999}
R.~Casalbuoni and R.~Gatto,
Phys.\ Lett.\ B {\bf 464}, 111 (1999) [hep-ph/9908227].

\bibitem{son:1999}
D.T.~Son and M.A.~Stephanov,
Phys.\ Rev.\ D {\bf 61}, 074012 (2000) [hep-ph/9910491] \newline
[Erratum-{\em ibid.} {\bf 62}, 059902 (2000), hep-ph/0004095].

\bibitem{bed:2001}
P.F.~Bedaque and T.~Sch\"afer,
Nucl.\ Phys.\ A {\bf 697}, 802 (2002) [hep-ph/0105150].

\bibitem{sch:2001}
T.~Sch\"afer,
Phys.\ Rev.\ D {\bf 65}, 074006 (2002) [hep-ph/0109052].

\bibitem{rho:2000}
M.~Rho, E.V.~Shuryak, A.~Wirzba and I.~Zahed,
Nucl.\ Phys.\ A {\bf 676}, 273 (2000) [hep-ph/0001104].

\bibitem{ban:1985}
M.~Bando, T.~Kugo, S.~Uehara, K.~Yamawaki and T.~Yanagida,
Phys.\ Rev.\ Lett.\  {\bf 54}, 1215 (1985).

\bibitem{ban:1987}
M.~Bando, T.~Kugo and K.~Yamawaki,
Phys.\ Rept.\  {\bf 164}, 217 (1988).

\bibitem{pis:1996}
R.D.~Pisarski and M.~Tytgat,
Phys.\ Rev.\ D {\bf 54}, 2989 (1996) [hep-ph/9604404].

\bibitem{Casalbuoni:2000jn}
R.~Casalbuoni, Z.~Duan and F.~Sannino,
Phys.\ Rev.\ D {\bf 63}, 114026 (2001) [hep-ph/0011394].

\bibitem{KSRF}
K.~Kawarabayashi and M.~Suzuki,
Phys.\ Rev.\ Lett.\  {\bf 16}, 255 (1966); \\
Riazuddin and Fayyazuddin,
Phys.\ Rev.\  {\bf 147}, 1071 (1966).

\bibitem{Jackson:2003dk}
A.D.~Jackson and F.~Sannino,
hep-ph/0308182.

\bibitem{man:2000}
C.~Manuel and M.H.G.~Tytgat,
Phys.\ Lett.\ B {\bf 501}, 200 (2001) [hep-ph/0010274].

\bibitem{har:2001}
M.~Harada and K.~Yamawaki,
Phys. Rev. Lett. {\bf 87}, 152001 (2001) [hep-ph/0105335].

\bibitem{kap:1991}
J.~Kapusta, P.~Lichard and D.~Seibert,
Phys.\ Rev.\ D {\bf 44} (1991) 2774 \newline
[Erratum-{\em ibid.}\ D {\bf 47} (1993) 4171].

\bibitem{song:1993}
C.~Song,
Phys.\ Rev.\ C {\bf 47}, 2861 (1993).

\bibitem{sho:2002a}
I.A.~Shovkovy and P.J.~Ellis,
Phys.\ Rev.\ C {\bf 66}, 015802 (2002) [hep-ph/0204132].

\bibitem{sho:2002b}
I.A.~Shovkovy and P.J.~Ellis,
Phys.\ Rev.\ C {\bf 67}, 048801 (2003) [hep-ph/0211049].

\bibitem{ouy:2003} R.~Ouyed, R.~Rapp and C.~Vogt, in preparation.

\bibitem{Rapp:1995fv}
R.~Rapp and J.~Wambach,
Phys.\ Lett.\ B {\bf 351}, 50 (1995) [nucl-th/9502023].

\end{thebibliography}
\end{document}